\documentclass[journal,12pt,onecolumn,draftclsnofoot,]{IEEEtran}
\usepackage{amsmath,amssymb,amsfonts,amsthm,epsfig}
\usepackage{color,graphicx,lscape,longtable,xr}
\usepackage[round]{natbib}
\usepackage{url}
\usepackage{geometry}
\usepackage{hyperref}
\usepackage{verbatim}
\usepackage{epstopdf}
\usepackage{subfigure}
\allowdisplaybreaks

\newcommand{\var}{\mbox{var}}
\renewcommand{\hat}{\widehat}

\usepackage{bm}
\usepackage{float}
\usepackage{multirow,color}
\usepackage{booktabs}
\usepackage{caption}
\usepackage{mathrsfs}
\usepackage{tikz}
\def\supw{\sup_{\w\in{\cal W}}}
\newcommand{\calW}{{\cal W}}

\def\bse{\begin{eqnarray*}}
\def\ese{\end{eqnarray*}}
\def\be{\begin{eqnarray}}
\def\ee{\end{eqnarray}}
\def\bsq{\begin{equation*}}
\def\esq{\end{equation*}}
\def\bq{\begin{equation}}
\def\eq{\end{equation}}

\def\wh{\widehat}
\def\wt{\widetilde}

\def\n{\nonumber}

\def\argmin{\mbox{argmin}}

\def\bSa{{\boldsymbol\Sigma}}

\def\bmu{{\boldsymbol\mu}}

\def\0{{\bf 0}}
\def\1{{\bf 1}}

\def\N{\mbox{ $\mathcal{N}$}}

\def\w{{\bf w}}
\def\X{{\bf X}}
\def\x{{\bf x}}

\def\bmu{{\boldsymbol \mu}}

\def\wh{\widehat}
\def\wt{\widetilde}

\def\sums{\sum\nolimits_{s=1}^{{S_n}}}

\def\sumi{\sum_{i=1}^n}

\def\sums{\sum_{s=1}^S}

\newcommand{\bit}{\begin{itemize}}
\newcommand{\eit}{\end{itemize}}

\newtheorem{Def}{Definition}
\newtheorem{Th}{\underline{\bf Theorem}}

\newtheorem{Assump}{\underline{Assumption}}

\def\boxit#1{\vbox{\hrule\hbox{\vrule\kern6pt  \vbox{\kern6pt#1\kern6pt}\kern6pt\vrule}\hrule}}
\def\bse{\begin{eqnarray*}}
\def\ese{\end{eqnarray*}}
\def\be{\begin{eqnarray}}
\def\ee{\end{eqnarray}}
\def\bsq{\begin{equation*}}
\def\esq{\end{equation*}}
\def\bq{\begin{equation}}
\def\eq{\end{equation}}

\def\wh{\widehat}

\def\wt{\widetilde}

\def\mR{\mathbb{R}}
\def\n{\nonumber}

\def\argmin{\mbox{argmin}}

\def\trans{^{\rm T}}

\def\bmu{\boldsymbol\mu}

\def\N{\mbox{ $\mathcal{N}$}}

\def\w{{\bf w}}

\def\Z{{\bf Z}}

\def\log{\hbox{log}}

\def\squarebox#1{\hbox to #1{\hfill\vbox to #1{\vfill}}}

\def\0{{\bf 0}}

\def\var{\hbox{var}}
\def\cov{\hbox{cov}}

\def\Pr{\rm p}
\def\wh{\widehat}
\def\wt{\widetilde}

\def\log{\hbox{log}}

\begin{document}
\title{
Frequentist Model Averaging for Global Fr\'{e}chet Regression}

\author{
\ Xingyu Yan\textsuperscript{a,b}
\ Xinyu Zhang\thanks{Correspondence address: xinyu@amss.ac.cn }\textsuperscript{c,b}
\ Peng Zhao\textsuperscript{a}\\
\textsuperscript{a}School of Mathematics and Statistics, \\
Jiangsu Normal University, Xuzhou, China \\
\textsuperscript{b}International Institute of Finance, \\
University of Science and Technology of China, Hefei, China \\
\textsuperscript{c}Academy of Mathematics and Systems Science, \\
Chinese Academy of Sciences, Beijing, China
}

\date{\today}
\maketitle

\def\shorttitle.arg{
Model averaging
}

\begin{abstract}
To consider model uncertainty in global Fr\'{e}chet regression and improve density response prediction, we propose a frequentist model averaging method. The weights are chosen by minimizing a cross-validation criterion based on Wasserstein distance. In the cases where all candidate models are misspecified, we prove that the corresponding model averaging estimator has asymptotic optimality, achieving the lowest possible Wasserstein distance. When there are correctly specified candidate models, we prove that our method asymptotically assigns all weights to the correctly specified models. Numerical results of extensive simulations and a real data analysis on intracerebral hemorrhage data strongly favour our method.
\noindent
\\[6pt]
\noindent{\bf Index Terms}:
Asymptotic optimality,
Cross-validation,
Fr\'{e}chet regression,
Model averaging,
Model uncertainty,
Wasserstein distance
\end{abstract}


\section{Introduction}

Data consisting of samples of
probability density functions are
increasingly prevalent
in various scientific fields, such as
biology, econometrics, and
medical science.
Examples include
population age and
mortality distributions across
different countries or regions
\citep{bigot2017geodesic,petersen2019frechet},
as well as the distributions of
functional magnetic resonance imaging (MRI) scans
in the brain
\citep{petersen2016functional}.
Despite the growing popularity
of probability density function data,
statistical methods
for analyzing such data are limited,
with only a few existing works available
\citep[e.g.,][]{petersen2021wasserstein,zemel2019frechet,
han2019additive,petersen2022modeling,chen2021wasserstein,
tucker2023variable,lin2023causal}.
The majority of current research
focuses on methods for
depicting
the association
between densities and Euclidean or non-Euclidean predictors
through estimated
conditional mean densities,
which are defined
as conditional Fr\'{e}chet means
under a suitable metric.
However, similar to the
traditional regression framework,
much of the
practical interest
in Fr\'{e}chet regression
applications lies in prediction,
rather than solely in the
inherent
density-predictor relationships.

In Fr\'{e}chet regression,
there exists model uncertainty
on which predictors practitioners
should use. Model selection is an
attempt to choose a single best model,
with the aim of improving prediction accuracy.
However, the selected model may suffer
from loss of some useful information contained
in other models \citep{bates1969combination}.
Furthermore, the results of model selection
can be unstable when there are minor changes in the data,
leading to inaccurate prediction
performance in practical applications \citep{yuan2005combining}.
Model averaging is an alternative approach
in dealing with model uncertainty
and improving prediction accuracy.
Instead of selecting a single model,
model averaging
combines candidate models by
assigning different weights to
each candidate model. This
approach often reduce
the prediction risk in regression
estimation because multiple models
provide a type of insurance against
the possible poor performance of a singly selected model
\citep{leung2006information}.

There are two mainstream approaches
to model averaging:
one from the Bayesian perspective and
the other on the frequentist basis.
Bayesian model averaging (BMA) has long
been a popular statistical strategy;
see \cite{hoeting1999bayesian,raftery1997bayesian},
and the references therein,
but the choice of appropriate priors in BMA
often remains unclear and relies on experiential
knowledge.
In recent years, frequentist model
averaging (FMA)
has also attracted
abundant attention
as it emerges as an impressive forecasting
device
in many applications,
such as, meteorology, social sciences,
finance and so on.
The FMA method
takes advantage of helpful
information of all candidate models
by assigning
heavier weights to stronger
candidate
models based on different selection criteria.
Currently there exists a large body of literature written
on this subject
\citep[e.g.,][]{buckland1997model,yuan2005combining,
hansen2007least,liang2011optimal,lu2015jackknife,
zhang2019parsimonious}.
As the data structures become more complex,
\cite{zhang2013model}
employ the jackknife criterion to choose the
optimal model weights
under dependent data.
\cite{gao2016model}
investigate a FMA
method based on the leave-subject-out
cross-validation
under a longitudinal data setting.
\cite{feng2022model} provide
a nonlinear model averaging framework
and suggest a new weight-choosing criterion.
\cite{liu2020model}
study the optimal model averaging in time series models
to improve the practical predictive
performance.
For multi-category responses,
\cite{li2022adaboost}
combine a semiparametric model
averaging  approach
with AdaBoost algorithm to obtain
more accurate estimations
of class probabilities.

All previous research findings
are based on averaging various
Euclidean regression models.
How to
extend the concept of model averaging
to Fr\'{e}chet regression framework
is not yet clear.
This limitation apparently hinders
the application of model averaging methods
in contemporary data analysis.
To address the need for
predictive studies within the
Fr\'{e}chet regression framework,
to our knowledge,
we for the first time propose a
model averaging approach
for Fr\'{e}chet regression problems.
This extension is challenging because
any linear combined predictions do not reside in
general metric space, such as manifold
and spherical response.
To avoid such issues,
we simplify the problem by
concerning the
Fr\'{e}chet regression for
probability density functions
with
the Wasserstein distance.
Methodologically,
we consider a global
Fr\'{e}chet regression setup
with density types of response
and
develop a frequentist model
averaging method
that combines
Fr\'{e}chet estimators
of each candidate model.
Furthermore,
a selection criterion
by $K$-fold cross-validation based on Wasserstein distance
is devised to appropriately choose
the weights of candidate models.
This
strategy aims to
assist researchers in
achieving improved practical
predictive performance.
Theoretically,
we first rigorously
prove that the proposed
averaging prediction using
$K$-fold cross-validation weights
is asymptotically optimal in
the sense of achieving the lowest
possible prediction risk, when all
candidate models are
misspecified.
Second,
when the model set includes
correctly specified models,
we establish that the proposed
approach asymptotically assigns all
weights to the correctly specified models,
i.e., the consistency of weights.
The proposed
$K$-fold cross-validation
model averaging
method
is intuitive and
easy for implementation.
Simulation studies and an
application on intracerebral hemorrhage data demonstrate
the advantages of the proposed
method.

The remaining part of the paper is organized as follows.
In Section \ref{sec:pre}, we briefly describe
the necessary concepts of the
Fr\'{e}chet regression model.
The main ideas for the proposed model averaging
approach are in Sections \ref{sec:method}
and \ref{sec:weight},
including
a detailed description about the
proposed modelling strategy
and the resulting prediction,
a weight choice criterion
based on minimizing the
Wasserstein distance
of the model averaging estimator.
Theoretical properties of the proposed
prediction and estimation involved
are presented in Section \ref{sec:asy}.
In Section \ref{sec:sim}, we conduct intensive
simulation studies to demonstrate how well
the proposed prediction works.
The simulation results show
the proposed method indeed leads
to more accurate predictions
than its alternatives.
In Section \ref{sec:real}, we present a
case study evidence on
intracerebral hemorrhage data,
to further illustrate the advantages of the
proposed method.
Finally, Section \ref{sec:dis} concludes
the paper with a short discussion.
All theoretical proofs are
left in the Appendix.

\section{A modelling strategy in
density response prediction}\label{sec:2}
\subsection{Preliminaries}\label{sec:pre}

Since the Fr\'{e}chet regression
introduced by \cite{petersen2019frechet}
is still relatively new
in statistics although
there are many applications
in other fields,
we provide a brief review
in this section before turning
to model averaging prediction for
Fr\'{e}chet regression with density response
in next subsection.

To facilitate the discussion,
let $(\Omega, d)$ be a metric space
equipped with a specific metric $d$,
and $\mR^p$ be the
$p$-dimensional Euclidean space.
For given metric space $\Omega$,
the seminal work of \cite{Frechet1948}
generalizes the
conventional
concepts of mean and variance
to the Fr\'{e}chet version as
\be
\mu_Y=\underset{\omega \in \Omega}{\argmin}~
\mathbb{E} \left\{d^2(Y, \omega)\right\},
\quad V_Y=\mathbb{E} \left\{d^2\left(Y, \mu_Y\right)\right\},
\label{eq:fre}
\ee
where $\mu_Y$ and $V_Y$ coincide with the
classical mean and variance when
$\Omega=\mR$.
Recall that, when $\Omega=\mR$,
the central role of classical regression is
to estimate the conditional expectation
\bse
m(\x)&=&\mathbb{E} (Y \mid \X=\x)
\\&=&\underset{y \in \mR}{\argmin}~
\mathbb{E}  \left\{(Y-y)^2 \mid \X=\x\right\} .
\ese
Replacing the Euclidean distance with
the intrinsic metric $d$ of $\Omega$,
\cite{petersen2019frechet}
define the general concepts of
Fr\'{e}chet regression function
of $Y$ given $\X=\x$ as follows
\be
m_{\oplus}(\x)&=&\underset{\omega \in \Omega}
{\operatorname{argmin}} M_{\oplus}(\x, \omega),
\label{eq:model0}
\ee
where
$
M_{\oplus}(\x, \omega)
=
\mathbb{E}
\left\{d^2(Y, \omega) \mid \X=\x\right\}.
$
Thus, the definitions of
\eqref{eq:fre} and
\eqref{eq:model0}
can be interpreted as
marginal and conditional Fr\'{e}chet
means, respectively.
Therefore, Fr\'{e}chet regression
aims at capturing the
relationship between
response $Y \in \Omega$
and predictors
$\X \in \mR^p$
by using conditional
Fr\'{e}chet
means.
In the paper,
we focus on the following
global Fr\'{e}chet regression,
which
is a generalization of
standard multiple linear regression
to the Fr\'{e}chet version.

\begin{Def}\label{def:1}
\citep{petersen2019frechet}
The global Fr\'{e}chet
regression model is characterized by,
for any $ \x \in \mR^p$,
\bse
m_{\oplus}(\x)
&=&
\arg\min _{\omega \in \Omega}
\mathbb{E}  \left\{
s(\X,\x)
d^2(Y, \omega)\right\},
\ese
where
$
s(\X,\x)
=1+(\x-\boldsymbol{\mu})\trans
\bSa^{-1}(\X-\boldsymbol{\mu}),
$
${\boldsymbol\mu}=\mathbb{E} (\X)$ and
$\boldsymbol\Sigma=\var(\X)$
are the traditional mean and
covariance matrix of $\X$.
\end{Def}

The global Fr\'{e}chet regression model
is emphasized as a ``global" regression because
it works effectively on arbitrary metric
spaces without requiring any tuning
parameters or local smoothing techniques.
Moreover,
from the Definition \ref{def:1},
the global
Fr\'{e}chet regression model
is
applicable for multiple predictors
and
does not
feature regression coefficients.
This lack of parameters makes
it a major challenge
to extend existing model averaging methods
for conventional regression to global Fr\'{e}chet framework.
Recent developments of Fr\'{e}chet regression
under other settings may be found in, for example,
\cite{dubey2020frechet,ghodrati2022distribution,
zhang2021dimension,lin2023causal}
and references therein.

\subsection{Models and weighted average prediction}\label{sec:method}

Adopting the framework of
Fr\'{e}chet regression for
probability density functions
with Euclidean predictors,
we consider
a particular
global Wasserstein-Fr\'{e}chet regression.
For convenience,
let
$Y_1^{-1}$ and $Y_2^{-1}$ denote
the quantile functions corresponding
to $Y_1$ and $Y_2$, respectively,
and let
$d_W^2(Y_1, Y_2)$
denote the
$\mathcal{L}^2$-Wasserstein distance
\citep{petersen2021wasserstein},
which is defined as
$
d_W(Y_1, Y_2)
=[
\int_0^1\{Y_1^{-1}(t)-Y_2^{-1}(t)\}^2 ~d t
]^{1 / 2}.
$
Assume that the space $\Omega$
is a set of probability density functions
equipped with the Wasserstein distance
and takes the form of a
weighted Fr\'{e}chet mean
\bse
m_{\oplus}(\x)
&=& \underset{\omega \in \Omega}{\arg\min}~
\mathbb{E}
\left\{
s(\X,\x)
d^2_W(Y, \omega)\right\}.
\ese
Here, the weight refers to $s(\X,\x)$, as provided
in Definition \ref{def:1},
and $\Omega$ is referred to
as the Wasserstein space.
Note that $m_{\oplus}(\x)$ denotes
the Fr\'{e}chet regression function
or conditional Wasserstein means,
and we assume the existence
and uniqueness of these quantities
throughout this paper.

In practice,
it is often the case that
there are typically just a few relevant variables
among the predictors $\X\in\mR^p$
that have been recorded.
Using too many or too few predictors
can lead to biased fitting and inaccurate model predictions.
Consequently, there exists model uncertainty
in the utilization of predictors,
making it
necessary to
employ model selection
or model averaging method to
reduce the prediction risk.
Model selection is sometimes unstable
because even a slight change
in the data can lead to a
significant change in the model choice results.
Thus,
an alternative sensible approach would
be applying the model averaging idea
to construct the prediction.
To improve the predictive performance
of global Wasserstein Fr\'{e}chet regression,
we develop a frequentist model
averaging estimation of the
conditional Wasserstein means.
Specifically,
consider a sequence of candidate models
$s=1,\ldots, S$, and the $s$th
candidate model uses the following
global Fr\'{e}chet regression function
\bse
&&
m_{\oplus}(\x^{(s)})\\
&=& \underset{\omega \in \Omega}{\arg\min}~
\mathbb{E} \left[\left\{
1+(\x^{(s)}-\bmu_{(s)})\trans
{\bSa}_{(s)}^{-1}
(\X^{(s)}-\bmu_{(s)})\right\}
d_W^2(Y, \omega)\right],
\ese
where
$\x^{(s)}\in\mR^{p_s}$ denotes
an
interested future
observation in the domain
of the $s$th model.
$p_s$ is the dimension of the $s$th model.
$\boldsymbol{\mu}^{(s)}=\mathbb{E}(\X^{(s)})$
and ${\bSa}^{(s)}=\cov(\X^{(s)})$
for
$s=1,\ldots,S$.

Let $F$ be the joint distribution of
$(\X, Y)$
defined on $\mR^p \times \Omega$.
Given an independent and identically
distributed (i.i.d.) sample
$\mathcal{D}_n=\left\{\left(\X_1, Y_1\right),
\ldots,\left(\X_n, Y_n\right)\right\}$
with $\left(\X_i, Y_i\right) \sim F$.
In practice,
the sample version of
the $s$th candidate model
$m_{\oplus}(\x^{(s)})$
is defined as
\be
&& \wh{m}_{\oplus}(\x^{(s)}) \n\\
&=& \underset{\omega \in \Omega}{\arg\min}
\sumi
\left\{
1+(\x^{(s)}-\overline{\X}^{(s)})\trans
\wh{\bSa}^{-1}_{(s)}
\left(\X_i^{(s)}-\overline{\X}^{(s)}\right)
\right\} d^2_W\left(Y_i, \omega\right),
\label{eq:fm}
\ee
where
$\overline{\X}^{(s)}=n^{-1} \sumi \X_i^{(s)}$
and
$\wh{\bSa}_{(s)}=
n^{-1} \sumi(\X_i^{(s)}-\overline{\X}^{(s)})
(\X_i^{(s)}-\overline{\X}^{(s)})\trans$
denote the sample mean and
sample covariance matrix
in the $s$th model,
respectively,
and $\X_i^{(s)}\in\mR^{p_s}$
represents the predictors used in the $s$th model
for $i=1,\ldots,n$.
Detailed optimization algorithm of these estimators
is given in Section 6.1 of
\cite{petersen2019frechet}.

Now,
let $\w=\left(w_1, \ldots, w_S\right)\trans$
be a weight vector with $w_s \geq 0$
and $\sum_{s=1}^S w_s=1$.
That is, the weight vector
$\mathbf{w}$
belongs to the continuous set
$\mathcal{W}=\{\w \in[0,1]^S: \sums w_s=1\}$.
Combining all possible predicted values of
$\wh{m}_{\oplus}(\x^{(s)})$
$(s=1,\ldots,S)$,
we construct an averaging
global Fr\'{e}chet
regression estimator
as
\be
\wh{m}_{\oplus}(\w)
&=&
\sums w_s\wh{m}_{\oplus}(\x^{(s)}).
\label{eq:MAest}
\ee
Note that
the constructed
averaging estimator is obviously
also in
Wasserstein space $\Omega$
since the Wasserstein
distance between the two distributions
is actually the Eucliden distance between
their quantile
points.
In the following development,
we will determine the optimal weights
and
design a procedure to predict the
conditional Wasserstein means.
\subsection{Weight choice criterion}\label{sec:weight}

As we can see,
the weights $w_s$'s in \eqref{eq:MAest} play
a key role in the success of the
model averaging prediction.
We use $K$-fold cross-validation
to choose the weights.
This section describes how
to calculate the $K$-fold
cross-validation criterion and construct an
averaging prediction with data-driven weights in detail.
For ease of presentation,
the introduced procedure is
summarized
by the following steps.

\noindent{\bf Step 1:}
Divide the data set into $K$ groups with
$2 \leq K \leq n$, so that there are $J=n / K$
observations in each group.
For simplicity of expression, we assume
that $J$ is an integer.

\noindent{\bf Step 2:}
For $k=1, \ldots, K$,
calculate the prediction for
an observation
at any $\x$
within
the $k$th group for each model.
That is, for $s=1, \ldots, S$,
we calculate
the prediction of $Y_{(k-1)J+j}$ by
\bse
&& \wh{m}_{\oplus}^{[-k]}
(\x_{(k-1)J+j}^{(s)}) \\
&=& \underset{\omega \in \Omega}{\arg\min}
\sum_{\ell\notin\{(k-1)J+1,\ldots,kJ\}}
\left[1+
\left\{\x_{(k-1)J+j}^{(s)}
-\overline{\X}_{[-k]}^{(s)}\right\}\trans
\wh{\bSa}_{[-k],(s)}^{-1}
\left\{\X_{\ell}^{(s)}-\overline{\X}_{[-k]}^{(s)}\right\}
\right] d^2_W\left(Y_{\ell}, \omega\right),
\ese
for $j=1,\ldots, J$,
where
the subscript $(k-1)J+j$
indicates the observations in the $k$th group
and subscript $\ell$ belongs to the
remaining $n-J$
observations excluding the $k$th group
from the data set.
The sample mean and covariance matrix
without the $k$th group
are calculated by
$\overline{\X}^{(s)}_{[-k]}
=(n-J)^{-1}\sum_{\ell\notin\{(k-1)J+1,\ldots,kJ\}}
 \X_{\ell}^{(s)}$
and
$\wh{\bSa}_{[-k],(s)}=
(n-J)^{-1}\sum_{\ell\notin\{(k-1)J+1,\ldots,kJ\}}
(\X_{\ell}^{(s)}-\overline{\X}^{(s)}_{[-k]})
(\X_{\ell}^{(s)}-\overline{\X}^{(s)}_{[-k]})\trans$,
respectively.

Therefore,
our $K$-fold
cross-validation criterion
is
constructed as follows
\be
\text{CV}_K(\w)&=&
\sum_{k=1}^K\sum_{j=1}^J
d_W^2\left\{Y_{(k-1)J+j},
\wh{m}_{\oplus,(k-1)J+j}^{[-k]}(\w)
\right\},
\label{eq:cv}
\ee
where
\bse
\wh{m}_{\oplus,(k-1)J+j}^{[-k]}(\w)
&=&\sums w_s
\wh{m}_{\oplus}^{[-k]}(\x_{(k-1)J+j}^{(s)}).
\ese

\noindent{\bf Step 3:}
Select the model weight vector by minimizing the
$K$-fold cross-validation criterion
\be\label{eq:opt}
\wh\w
&=&
\underset{\w\in \mathcal{W}}
{\argmin} \
\text{CV}_K(\mathbf{w}),
\ee
and by formula of \eqref{eq:MAest}
an averaging prediction
for the conditional Wasserstein means
from these $S$ models
is
as follows
\be
\wh m_{\oplus}(\wh\w)
&=&
\sums \wh w_s
\wh{m}_{\oplus}
(\x^{(s)}).
\label{eq:MAwest}
\ee

As demonstrated by
\cite{arlot2016choice},
when the fold $K$ is set to $5$ or $10$,
the performance of $K$-fold cross-validation
can be close to be optimal.
We adopt $K=10$ for ease of computation
throughout this paper.
The above optimization \eqref{eq:opt}
can also be solved easily and rapidly,
as in traditional model averaging
it can be implemented with many
existing optimize functions in R or Matlab.
In our numerical studies,
we use the ``fmincon" function in R language.

\section{Theoretical properties}\label{sec:asy}
In this section, we present the asymptotic
properties of the proposed averaging
prediction.
All limiting processes
discussed here and throughout the text
are with respect to
$n\rightarrow\infty$.
To facilitate the theoretical analysis,
we will use the notation
$Y^{-1}(t)$ and
$Q_{\oplus}(\x,t)$ to denote
the quantile value of the probability
density function $Y$
and conditional Wasserstein means
$m_{\oplus}(\x)$ at argument
$t\in[0,1]$, respectively.
By the Wasserstein distance,
we let the risk function be
\be\label{eq:rw}
r(\w)&=& \mathbb{E}
 d_W^2\{m_{\oplus}(\x_0),\wh m_{\oplus,0}(\w)\}
\n\\&=&
\mathbb{E}
\int_0^1 \left\{
Q_{\oplus}(\x_0,t)- \wh Q_{\oplus,0}(\w,t)
\right\}^2dt,
\ee
where $\wh m_{\oplus,0}(\w)=
\sums w_s\wh{m}_{\oplus}(\x_0^{(s)})$
with $\x_0^{(s)}$ being a new predictor vector,
and the corresponding
random
quantile function is
$\wh Q_{\oplus,0}(\w,t)=
\sums w_s\wh{Q}_{\oplus}(\x_0^{(s)},t)$.
Then, we establish the theoretical properties of the procedure
by examining its performance in terms of minimizing $r(\w)$.
We further introduce some notations before we
present the optimality of the selected model weights.
Define the average leave-one-out prediction to be
\bse
\wh{m}_{\oplus,i}^{[-i]}(\w)
&=&\sums w_s
\wh{m}_{\oplus}^{[-i]}(\x_i^{(s)}),
\ese
where
$\wh{m}_{\oplus}^{[-i]}(\x_i^{(s)})$
is the leave-one-out prediction of
$Y_i$
in the $s$th model,
and similarly for
$\wh{Q}_{\oplus,i}^{[-i]}(\w,t)
=\sums w_s
\wh{Q}_{\oplus}^{[-i]}(\x_i^{(s)},t)$
with $\wh{Q}_{\oplus}^{[-i]}(\x_i^{(s)},t)$
being the prediction of $Y_i^{-1}(t)$.
Let
\be\label{eq:Rw}
R(\w)
&=&\mathbb{E}
\frac{1}{n}\sumi
d_W^2\{m_{\oplus}(\x_i),\wh{m}_{\oplus,i}^{[-i]}(\w)\}
\n\\&=&\mathbb{E}
\frac{1}{n}\sumi
\int_0^1 \left\{
Q_{\oplus}(\x_i,t)-\wh{Q}_{\oplus,i}^{[-i]}(\w,t)
\right\}^2dt,
\ee
and
\be\label{eq:Rbar}
\bar{R}(\w)
&=&\mathbb{E}
\frac{1}{n}\sumi
d_W^2\{m_{\oplus}(\x_i),\wh{m}_{\oplus,i}(\w)\}
\n\\&=&\mathbb{E}
\frac{1}{n}\sumi
\int_0^1 \left\{
Q_{\oplus}(\x_i,t)-\wh{Q}_{\oplus,i}(\w,t)
\right\}^2dt.
\ee
The risk defined in
\eqref{eq:rw} is
associated with
the leave-one-out prediction risk $R(\w)$
in \eqref{eq:Rw}, and \eqref{eq:Rbar}
can be viewed as
sample version of \eqref{eq:rw}.

Finally,
suppose that,
for any fixed $\x^{(s)}\in\mR^p$,
there exists a limiting
function
$m_{\oplus}^\ast(\x^{(s)})$
for $\wh{m}_{\oplus}(\x^{(s)})$.
Then,
we introduce notation
associated with the limiting function as follows.
Let
\be\label{eq:Rstar}
R^\ast(\w)
&=&\frac{1}{n}\sumi
d_W^2\{m_{\oplus}(\x_i),m^\ast_{\oplus,i}(\w)\}
\n\\&=&
\frac{1}{n}\sumi
\int_0^1 \left\{
Q_{\oplus}(\x_i,t)-Q^\ast_{\oplus,i}(\w,t)
\right\}^2dt,
\ee
where
$m^\ast_{\oplus,i}(\w)=\sums
w_s m_{\oplus}^\ast(\x_i^{(s)}),
$
$Q^\ast_{\oplus,i}(\w,t)=\sums
w_s Q_{\oplus}^\ast(\x_i^{(s)},t),
$
and $\xi_n=n\inf_{w\in\cal W}R^\ast(\w)$.
\subsection{Asymptotic optimality}

To obtain the asymptotic optimality of
the proposed estimator, we impose the
following assumptions.

\begin{Assump}\label{C:1}
Assume that
the limiting
function $m_{\oplus}^\ast(\x^{(s)})$
satisfies
$
d_W\{ \wh{m}_{\oplus}(\x^{(s)}) ,
m_{\oplus}^\ast(\x^{(s)})\}=O_p(c_n)$,
where $c_n=n^{-1/(2(\beta-1))}$ with $\beta>1$.
\end{Assump}

\begin{Assump}\label{C:2}
$Q_{\oplus}(\x_i,t)=O_p(1)$,
$Q_{\oplus}^\ast(\x_i^{(s)},t)=O_p(1)$
and $Y_i^{-1}(t)=O_p(1)$
hold uniformly for $t\in[0,1]$ and
$i\in\{0,1,\ldots,n\}$.
\end{Assump}

\begin{Assump}\label{C:3}
$\xi_n^{-1}n c_n=o_p(1)$
and $\xi_n^{-1}n^{1/2}S^2=o_p(1)$.
\end{Assump}

\begin{Assump}\label{C:Jfold}
$
\wh{Q}_{\oplus}(\x_{(k-1)J+j}^{(s)},t)-
\wh{Q}_{\oplus}^{[-k]}(\x_{(k-1)J+j}^{(s)},t)
=o_p(n^{-1}\xi_n),
$
$\wh{Q}_{\oplus}(\x_0^{(s)},t)
-\wh{Q}_{\oplus}^{[-1]}(\x_0^{(s)},t)=o_p(1),
$
and
$\wh{Q}_{\oplus}^{[-k]}(\x_{(k-1)J+j}^{(s)},t)=O_p(1)$
hold uniformly for $\w\in\calW$,
$t\in[0,1]$,
$j\in\{1,\ldots,J\}$
and $k\in\{1,\ldots,K\}$.
\end{Assump}

Assumption \ref{C:1} is taken from Theorem $2$ of
\cite{petersen2019frechet}.
This assumption ensures
that the estimator
$\wh{m}_{\oplus}(\x^{(s)})$
in each candidate model has a limit
$m_{\oplus}^\ast(\x^{(s)})$,
where $m_{\oplus}^\ast(\x^{(s)})$
might be interpreted as a pseudo-true value.
Similar assumption is commonly used
to analyze the asymptotic properties of the
model averaging estimator
in context of model averaging fields.
Assumption \ref{C:2}
quantifies the order of some involved terms,
which is used to
exclude some
pathological cases in which the limiting
value explodes.
Analogous requirement can be found in
\cite{panaretos2016amplitude,petersen2021wasserstein}.
The two conditions in Assumption \ref{C:3}
put bounds on the
order of prediction risk relative to the sample size
and series $c_n$, respectively.
It requires that $\xi_n$ grows at a rate no slower
than that of $nc_n$ and $n^{1/2}$.
When $\beta=2$,
the two conditions are identical since
$c_n=n^{-1/2}$.
That is $\xi_n$ grows at a rate
no slower than $n^{1/2}$,
which implies
that all candidate models are misspecified.
It is a common assumption in literature,
such as \cite{zhang2016optimal,zhang2023model}.
To further understand this condition,
supposing that $s_0$th model is correctly specified,
we have $m^\ast_{\oplus}(\x^{(s_0)})=m_{\oplus}(\x)$
for any $\x$. Then,
it follows that
\bse
\inf_{w\in\cal W}R^\ast(\w)
&=&\inf_{w\in\cal W}\frac{1}{n}\sumi
d_W^2\{m_{\oplus}(\x_i),m^\ast_{\oplus,i}(\w)\}
\\&\le&
\frac{1}{n}\sumi
d_W^2\{m_{\oplus}(\x_i),m^\ast_{\oplus}(\x_i^{(s_0)})\}
\\&=&0,
\ese
and thus $\xi_n=0$. This implies that Assumption \ref{C:3} is violated.
Therefore, if one of the candidate models is
correctly specified, then Assumption \ref{C:3}
does not hold.
Assumption \ref{C:Jfold} is an
intuitive result
for each candidate model
in general.
The first part
essentially means that
the difference between
$\wh{Q}_{\oplus}(\x_{(k-1)J+j}^{(s)},t)$
and the leave-$K$-out prediction
$\wh{Q}_{\oplus}^{[-k]}(\x_{(k-1)J+j}^{(s)},t)$
decreases with sufficient speed; the second part
requires that $\wh{Q}_{\oplus}(\x_0^{(s)},t)$
and $\wh{Q}_{\oplus}^{[-1]}(\x_0^{(s)},t)$
should be very close, which is also
reasonable as $n\rightarrow\infty$.
The aforementioned assumptions are often
used in the model selection and model
averaging literature. More similar detailed
discussions can be found in
\cite{ando2014model,zhang2018functional}
and references therein.

\begin{Th}\label{th:1}
Under Assumption \ref{C:1}-\ref{C:Jfold}, we have
\bse
\frac{r(\wh{\w})}
{\inf _{\w \in \mathcal{W}} r(\w)}
&\overset{\Pr}\rightarrow& 1,
\ese
where $\overset{\Pr}\rightarrow$
denotes convergence in probability.
\end{Th}

The proof of Theorem \ref{th:1}
is provided in the Appendix.
Theorem \ref{th:1} indicates that
the weight vector $\wh\w$ by the $K$-fold
cross-validation $\text{CV}_K(\w)$ based on
Wasserstein distance is asymptotically optimal
among all feasible weight vector choices.

\subsection{Consistency of weights}

In this section, we will demonstrate that
when there are some correctly specified models,
and the sample size is sufficiently large,
our approach can successfully identify
all these correctly specified models and
reduce the weights for
the misspecified models to zeros.
This conclusion corresponds to the
consistency property
in context of
model selection.

Specifically,
let $\cal{C}$ be the subset of
$\{1, \ldots, S\}$
that contains all the indices of the correctly
specified models,
and
$\cal{W}_S=
\{\w\in\cal{W}:\ $$\sum_{s\notin\cal{C}} w_s=1\}$
be the subset of $\mathcal{W}$ that
assigns all weights to the misspecified models.
We need
the following additional assumption.

\begin{Assump}\label{C:weight}
Assume that
$\inf_{\w\in\cal{W}_S}R^\ast(\w)\ge c$
for some constant $c>0$.
\end{Assump}

By Assumption \ref{C:weight}, we can
obtain that
$
c_n\{\inf_{\w\in\cal{W}_S}R^\ast(\w)\}^{-1}
=o_p(1),
$
which is equivalent to the first part
of Assumption \ref{C:3}, when subset $\cal{C}$ is empty, that is,
all candidate models are misspecified.
This result is commonly used
to ensure
consistency of weights
in traditional model averaging framework.
See, for example,
\cite{yu2022unified}.

\begin{Th}\label{th:2}
If Assumptions \ref{C:1}, \ref{C:2},
\ref{C:Jfold} and \ref{C:weight}
are satisfied, then
we have
\bse
\sum_{s \in \mathcal{C}} \wh{w}_s
&\overset{\Pr}\rightarrow& 1,
\ese
where $\wh w_s$ is the $s$th entry of $\wh\w$.
\end{Th}

The proof of Theorem \ref{th:2} is given in the Appendix.
Theorem \ref{th:2} is a kind of model selection
consistency in context of model averaging framework,
in which the method will automatically
exclude the misspecified models.
This result indicates
that
when the model set includes correctly specified models
and the sample size
is sufficiently large,
the proposed $K$-fold
cross-validation successfully assigns all
weights to the correctly specified models.
Theorem \ref{th:1} and \ref{th:2}
guarantee that our model averaging estimator
achieves optimal performance
in theory across various practical scenarios.

\section{Simulation studies}\label{sec:sim}
\subsection{Alternative methods}

In this section,
we conduct simulation experiments to
demonstrate the finite sample performance
of our
$K$-fold cross-validation
model averaging method, KCVMA.
We compare it with the AIC- and BIC-based model selection
and averaging estimators,
as well as a ridge-type regularization
model selection approach for
global F\'{r}echet
regression
\citep{tucker2023variable}.

For the
$s$th candidate model,
let $\wh\sigma^2_s=
n^{-1}\sumi
[d_W \{Y_i, \wh{m}_{\oplus}(\x_i^{(s)})\}]^2
$
denote the estimated residual
with conventional mean squared error
replaced by
Wassertein distance.
Then,
\bse
\mathrm{AIC}_s&=&
\log\left( \wh\sigma^2_s
\right)+ \frac{2p_s}{n},
\ese
and
\bse
\mathrm{BIC}_s&=&
\log\left(
\wh\sigma^2_s
\right)+ p_s\times\frac{\log(n)}{n},
\ese
where $p_s$ is the number of
parameters
in $s$th candidate model.
The above two criteria each
select a model that corresponds
to the smallest of their
respective scores as usual.
Further,
similar to \cite{buckland1997model},
two weight choices
for model averaging based on the
smoothed-version of the AIC and BIC
are defined as follows
\bse
\quad \mathrm{sAIC}_s=\exp \left(-\mathrm{AIC}_s / 2\right)
/ \sum_{\ell=1}^S \exp \left(-\mathrm{AIC}_{\ell} / 2\right),
\ese
and
\bse
\mathrm{sBIC}_s=\exp \left(-\mathrm{BIC}_s / 2\right)
/ \sum_{\ell=1}^S \exp \left(-\mathrm{BIC}_{\ell} / 2\right).
\ese
Due to its ease of use, the sAIC and sBIC
weight choice methods have been
used extensively in the traditional FMA literature,
see, such as, \cite{wan2010least,wang2012model}.

\subsection{Simulation designs}

For the sake of fair comparison,
the following simulation setting is taken by
\cite{tucker2023variable}.
Specifically,
the correlated scalar predictors
$X_j \sim \mathcal{U}(-1,1), j=1,2, \ldots, p$,
are generated in two steps:
(1)
$\Z=(Z_1, Z_2, \ldots, Z_p)\trans$
multivariate Gaussian with
$E\left(Z_j\right)=0$
and $\cov(Z_j, Z_{j^{\prime}})
=\rho^{|j-j^{\prime}|}$
for $j,j^{\prime}=1,\ldots,p$;
(2)
$X_j=2 \Phi\left(Z_j\right)-1$
for $j=1, \ldots, p$,
where $p=10$, $\rho=0.5$,
and $\Phi$
is the standard normal
distribution function.
The Fr\'{e}chet regression
function is given by
\bse
m_{\oplus}(\mathbf{x})
&=&\mathbb{E}\{Y(\cdot) \mid \X=\x\} \\
&=&\mu_0+
\beta\left(x_4+x_8\right)
+\left(\sigma_0+\gamma x_1\right)
\Phi^{-1}(\cdot).
\ese
Conditional on $\X$, the random response $Y$
is generated by adding noise as follows:
$Y=\mu+\sigma \Phi^{-1}$ with
\bse
\mu \mid \X &\sim& \N
\left(\mu_0+\beta\left(X_4+X_8\right), v_1\right), \\
\sigma \mid \X &\sim&
\operatorname{Gamma}
\{(\sigma_0+\gamma X_1)^2 / v_2, v_2 /
(\sigma_0+\gamma X_1)\}.
\ese
being independently sampled.
Then, the important
predictors are $X_1$, $X_4$,
and $X_8$. The additional parameters are
set as $\mu_0=0$, $\sigma_0=3$,
$\beta=3/4$,
$\gamma=1$, $v_1=1$, and $v_2=0.5$.

The choice of candidate models is based on the method
of choosing individual ridge regularization parameters,
as proposed by \cite{tucker2023variable}.
The approach employs different
regularization parameters for each predictor component,
resulting in varying sets of
relevant variables selected among
the predictors for each regularization parameter.
Subsequently, we view these distinct sets of
relevant variables as candidate models.
More specific,
the regularization parameter
is denoted by $\lambda_j(\tau)$ with
a prespecified grid for $\tau$,
say $\tau\in\{\tau_1\le\tau_2\le\ldots\tau_K\}$.
Once we estimate
$\hat{\lambda}_j(\tau_k)$
for each of $\tau_k$,
we can estimate the relevant set of predictors as
$\hat{\mathcal{I}}(\tau_k)=
\{j: \hat{\lambda}_j(\tau_k)>0\}
\subset\{1, \ldots, p\}$.
These different sets among
$\{\hat{\mathcal{I}}(\tau_k),k=1,\ldots,K\}$ are seen
as our candidate models.
For fair of comparison,
we completely follow the
settings recommended
by \cite{tucker2023variable},
i.e., the grid $\{0.5, 1, 3p-0.5,3p\}$.

We compute model averaging estimators of
$m_{\oplus}(\mathbf{x}) $
by
$
\wh m_{\oplus}(\wh\w)
$
from \eqref{eq:MAwest}.
Our evaluation of the performance of
estimators is based on
the following average
Wasserstein distance loss or risk
\bse
\text{Risk}&=&\frac{1}{T} \sum_{r=1}^T
d_W^2\left\{\wh m_{\oplus}^{(r)}(\wh\w),
m_{\oplus}(\x)\right\},
\ese
where $T$ is the times of replication
and
$\wh m_{\oplus}^{(r)}(\wh\w)$
denotes the estimator
in the $r$th replication.
The
sample sizes of
$n$ are set as $100,200,300$, respectively,
in each simulation repetition.
We conduct $T= 100$ replications.

Besides our proposed method,
we also fit the data
with seven available methods including
(a) two conventional model averaging approaches
(sAIC, sBIC) presented by above subsection,
and equal weight model averaging method (EW);
(b) ridge-type shrinkage model selection method (Ridge)
recently introduced by \cite{tucker2023variable},
and the AIC and BIC type approaches also provided in
above subsection;
(c) ordinary least square estimation
under full model (Full)
proposed by \cite{petersen2019frechet},
and the oracle method (Oracle),
i.e., the unpenalized estimator
obtained
when the process of data generation is known.
We are going to examine the risk
of the above eight methods in the above
simulated cases.

The results of the simulations
are presented in
Table \ref{tab:rho5R9}.
In all scenarios, it can be observed
that the prediction performance of the model averaging
is better than that of the model selection methods.
Moreover, with the increase in sample size,
the estimated performance also improves.
In particular,
the KCVMA dominates the other methods
under different sample size,
that is, except Oracle,
the KCVMA exhibits the best prediction
accuracy in terms of risk, while the
AIC or BIC often performs the worst.
On average, these findings
suggest that the KCVMA is
better suited for prediction
in terms of risk.
In addition,
to examine the effect of autocorrelation of predictors
on the predictive results,
we also consider the scenarios of $\rho=0.2,0.8$
for a more comprehensive comparison.
The simulation results, similar to those with
$\rho=0.5$, are omitted here.

In addition,
we analyze the behavior
of the sum of the weights
$\sum_{s \in \mathcal{C}} \wh{w}_s$
assigned to correct models
in Table \ref{tab:weight}.
Again, the sample size $n$
takes value from $\{100,200,300\}$
and
$T = 100$ replications are generated.
In each replication,
$\sum_{s \in \mathcal{C}} \wh{w}_s$
is calculated,
then we average this
value over all the
replications.
It is observed that the
sum of model weights is monotonically increasing
and generally converges to one as the sample size increases.
This phenomenon supports the theoretical result of
Theorem \ref{th:2}.

\section{Empirical Application}\label{sec:real}

In this section, we analyze a practical example
with distribution function as the responses
to further examine
the effectiveness
of our
proposed model averaging method.
The dataset considered here is the
intracerebral hemorrhage (ICH) data,
which contain response observations on
the head CT hematoma densities of
total of $393$
ICH anonymous subjects, recorded as
smoothed probability density functions.
The covariates include $4$
radiological variables
and $5$ clinical variables as predictors.
The clinical predictors are
age, weight, history of diabetes and
two variables indicating history of coagulopathy (Warfarin and AntiPt).
Radiological predictors contain
the logarithm of hematoma volume,
a continuous index of hematoma shape
(Shape),
presence of a shift in the midline of the brain,
and length of the interval between
stroke event and the CT scan (TimetoCT).
A more detailed description of the data
source can be found in
\cite{hevesi2018untreated}.

To evaluate the
prediction accuracy of
each method and make a comparison between different
methods, we randomly split the
dataset into
training set of size $n_{\text{train}}$
and testing set of size $n_{\text{test}}$.
We apply each method under comparison to the
training set to form the hematoma density predictions, and use
the testing set to compute the out-of-sample
prediction error of this method.
For comparison,
we consider eight model averaging
and model selection methods
(e.g., sAIC, sBIC, EW, FULL, Ridge, AIC, BIC
and our CV approach)
presented in simulation studies
in Section \ref{sec:sim}.
We assess
the utility
of the methods considered via
the squared prediction errors (SPE),
defined as
$
\text{SPE}=
\sum_{i=1}^{n_{\text{test}}}d_W^2(\wt Y_i,Y_i)
/{n_{\text{test}}} ,
$
where $\wt Y_i$
and $Y_i$
denote the predicted values and observed
values in the testing set, respectively.
To facilitate comparison,
we scale the
SPE by subtracting
the lowest SPE across the eight model
averaging and selection methods
from the original SPE.
We repeat the procedure of randomly dividing
the sample into training and test
samples $50$ times
and set the size $n_{\text{train}}$
of the training set to be
$100$, $200$ and $300$, respectively.

\begin{figure}[h!]
\centering
\begin{tabular}{cc}
\includegraphics[height=4.0in,width=5.0in]{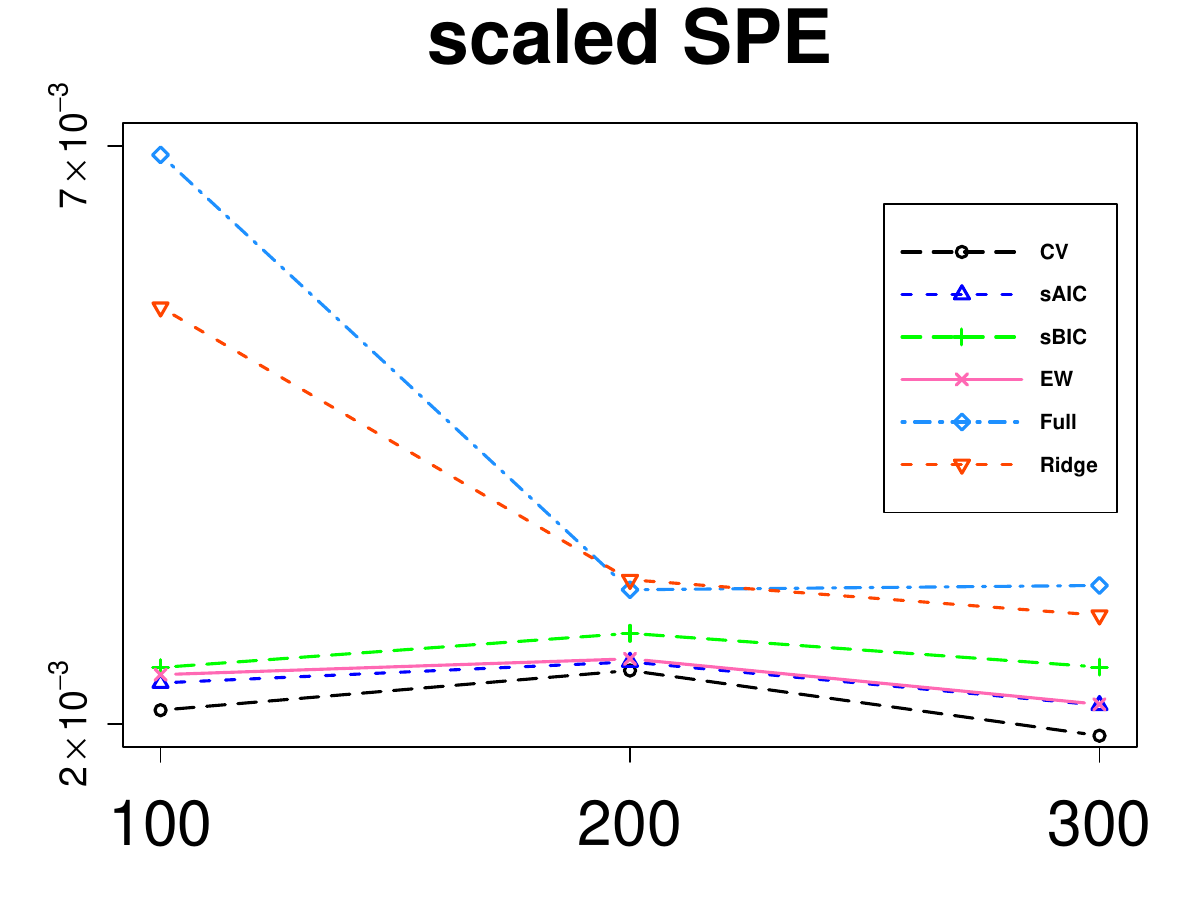}
\end{tabular}
\caption{
The scaled SPEs by various methods
in the intracerebral hemorrhage data analysis
for training sample size
$n_{\text{train}}=100,200,300$,respectively.
 }\label{fig:real}
\end{figure}

Figure \ref{fig:real}
presents
the scaled SPEs for
model averaging and model
selection methods
with training size $n_{\text{train}}=100,200,300$,
respectively.
For ease of presentation, we only
exhibit the results without
AIC and BIC, as these two approaches
have very poor performance.
The results show that our estimator
frequently produces the most accurate prediction
under all circumstances
and generally enjoys the smallest scaled SPE
among all estimators considered.
The above
numerical evidence
justifies the effectiveness of our method.
\section{Concluding Remark}\label{sec:dis}

In recent years, as data types are
becoming more complex, attention
has turned to regression in more abstract settings,
such as probability density function,
networks, manifolds and simplex-valued responses.
However,
similar to traditional regression setting,
model uncertainty in these abstract settings
is still inevitable.
The results of singly model selection approach
are unstable
and might
miss some useful information
contained in other models.
Moreover, in real-world problems,
the main focus of applications of
various regression is often on prediction
rather than solely on the relationships
between responses and predictors.
Therefore,
it becomes crucial to
address model uncertainty more properly
in the abstract regression framework
to make more reliable predictions.

We propose a model
averaging procedure to
improve prediction for
Fr\'{e}chet regression model where
density curves appear as
response objects.
A weight choice criterion
based on minimizing Wasserstein distance
of the model average estimator
is developed,
and the asymptotic optimality
of the resultant estimator
and consistency of weights
are established.
Additionally,
simulations and real data analysis
confirm that our proposed
approach outperforms other competitive
methods in prediction accuracy.
Although the proposed modelling strategy and
the resulting predictions are partially stimulated by a
particular dataset, apparently,
they are widely applicable for other
density response datasets
from many other disciplines.
In the future research,
we will extend
the proposed averaging method to high-dimensional
Fr\'{e}chet regression,
as well as other general types of
responses as mentioned above.
Understanding the asymptotic
results when the sample size
is limited and developing finite sample properties are also
very necessary in the future research.

\section*{Acknowledgments}
This work was supported by the
National Natural Science Foundation of China
(grant numbers: 71925007, 12101270 and 12325109).

\section*{Disclosure statement}
No potential conflict of interest was reported by the author(s).

\section*{Appendix}\label{sec:proof}
\setcounter{equation}{0}\renewcommand{\theequation}{A.\arabic{equation}}
\setcounter{subsection}{0}\renewcommand{\thesubsection}{A.\arabic{subsection}}

\noindent{
{\bf Proof of Theorem \ref{th:1}.}}
First of all, we notice that
\bse
\begin{aligned}
& \frac{r(\wh{\w})}{\inf _{\w \in \cal{W}} r(\w)}-1 \\
= & \sup _{\w \in \cal{W}}\left\{\frac{r(\wh{\w})}{r(\w)}-1\right\} \\
= & \sup _{\w \in \cal{W}}\left\{\frac{r(\wh{\w})}{R(\wh{\w})} \frac{R(\wh{\w})}{\bar{R}(\wh{\w})}
\frac{\bar{R}(\wh{\w})}{\bar{R}(\w)}
\frac{\bar{R}(\w)}{R(\w)} \frac{R(\w)}{r(\w)}-1\right\} \\
\leq & \sup _{\w \in \cal{W}} \frac{r(\w)}{R(\w)}
\sup _{\w \in \cal{W}} \frac{R(\w)}{\bar{R}(\w)}
\sup _{\w \in \cal{W}} \frac{\bar{R}(\wh{\w})}{\bar{R}(\w)}
\sup _{\w \in \cal{W}} \frac{\bar{R}(\w)}{R(\w)}
\sup _{\w \in \cal{W}} \frac{R(\w)}{r(\w)}-1 \\
= &
\sup _{\w \in\cal{W}} \frac{r(\w)}{R(\w)}
\sup _{\w \in \cal{W}} \frac{R(\w)}{\bar{R}(\w)}
\frac{\bar{R}(\wh{\w})}{\inf _{\w \in \cal{W}} \bar{R}(\w)}
\sup _{\w \in \cal{W}} \frac{\bar{R}(\w)}{R(\w)}
\sup _{\w \in \cal{W}} \frac{R(\w)}{r(\w)}-1.
\end{aligned}
\ese

Hence, to prove the Theorem \ref{th:1},
it suffices to show that,
as $n\rightarrow\infty$,
\be\label{eq:rR}
\supw\left|
\frac{r(\w)}{R(\w)}
-1\right|&\overset{\Pr}\rightarrow& 0
\ee
\be\label{eq:Th1:1}
\frac{\bar{R}(\wh{\w})}{\inf _{\w \in \cal{W}} \bar{R}(\w)}
&\overset{\Pr}\rightarrow& 1,
\ee
and
\be\label{eq:Th1:2}
\sup _{\w \in \cal{W}}
\left|\frac{\bar{R}(\w)}{R(\w)}-1\right|
&\overset{\Pr}\rightarrow& 0.
\ee

We now prove the above equations separately as follows.
We first note that
\bse
R(\w)
&=&\mathbb{E} \frac{1}{n}\sumi
d_W^2\{m_{\oplus}(\x_i),\wh{m}_{\oplus,i}^{[-i]}(\w)\}
\n\\&=&\mathbb{E}
\frac{1}{n}\sumi
\int_0^1 \left\{
Q_{\oplus}(\x_i,t)-\wh{Q}_{\oplus,i}^{[-i]}(\w,t)
\right\}^2dt
\\&=&
\mathbb{E}
\int_0^1 \left\{
Q_{\oplus}(\x_0,t)-\wh{Q}_{\oplus,0}^{[-1]}(\w,t)
\right\}^2dt
\\&=&
\mathbb{E}
\int_0^1 \left\{
Q_{\oplus}(\x_0,t)-\wh{Q}_{\oplus,0}(\w,t)
\right\}^2dt+
\mathbb{E}
\int_0^1 \left\{
\wh{Q}_{\oplus,0}(\w,t)
-\wh{Q}_{\oplus,0}^{[-1]}(\w,t)
\right\}^2dt
\\&&+2
\mathbb{E}\left[
\int_0^1 \left\{
Q_{\oplus}(\x_0,t)-\wh{Q}_{\oplus,0}(\w,t)
\right\}
\left\{
\wh{Q}_{\oplus,0}(\w,t)
-\wh{Q}_{\oplus,0}^{[-1]}(\w,t)
\right\}dt\right]
\\&=& r(\w)+
\mathbb{E}
\int_0^1 \left[
\sums w_s\left\{
\wh{Q}_{\oplus}(\x_0^{(s)},t)
  -\wh{Q}_{\oplus}^{[-1]}(\x_0^{(s)},t)
  \right\}
\right]^2dt
\\&&+2
\mathbb{E}\left(
\int_0^1 \left\{
Q_{\oplus}(\x_0,t)-\wh{Q}_{\oplus,0}(\w,t)
\right\}
\left[
\sums w_s\left\{
\wh{Q}_{\oplus}(\x_0^{(s)},t)
  -\wh{Q}_{\oplus}^{[-1]}(\x_0^{(s)},t)
  \right\}
\right]dt\right)
\\&=& r(\w)+ o_p(1),
\ese
uniformly for $\w\in\calW$,
where the last equality is due to
the Assumption \ref{C:2} and
Assumption \ref{C:Jfold}.
This implies \eqref{eq:rR}.

We next deal with equation \eqref{eq:Th1:1}.
Let
\bse
\text{CV}_K^\ast(\w)
&=&
\text{CV}_K(\w)
-\sumi d_W^2(m_{\oplus}(\x_i),Y_i)
\n\\&=&\text{CV}_K(\w)
-\sumi
\int_0^1 \left\{
Q_{\oplus}(\x_i,t)-Y_i^{-1}(t)
\right\}^2dt,
\ese
where the second term of right hand side of
above equation is unrelated to $\w$.
Therefore,
\bse
\wh\w&=&\underset{\w \in \cal{W}}{\arg\min}~
\text{CV}_K(\w)=
\underset{\w \in \cal{W}}{\arg\min}~
\text{CV}_K^\ast(\w).
\ese
Write
\bse
\text{CV}_K^\ast(\w)
&=&n\bar{R}(\w)+a_1(\w)+a_2(\w),
\ese
where
$a_1(\w)=\text{CV}_K^\ast(\w)-nR^\ast(\w)$
and $a_2(\w)=nR^\ast(\w)-n\bar{R}(\w)$.
By the proof of Theorem $1$ of \cite{wan2010least},
if we can show that
\be
\sup _{\w \in \cal{W}}
\frac{\left|a_2(\w)\right|}
{n R^\ast(\w)}&=& o_p(1),
\label{eq:Th1:3}
\ee
and
\be
\sup _{\w \in \cal{W}}
\frac{\left|a_1(\w)\right|}{n R^\ast(\w)} &=&  o_p(1) ,
\label{eq:Th1:4}
\ee
then \eqref{eq:Th1:1} can be established.
So, we next prove the equation \eqref{eq:Th1:3}.
Recall that
$\xi_n=n\inf_{w\in\cal W}R^\ast(\w)$,
and notice that
\be
&&
\supw
\frac{\left|a_2(\w)\right|}
{n R^\ast(\w)}\n\\
&\le& \xi_n^{-1}\supw\left|a_2(\w)\right|\n\\
&=&\xi_n^{-1}\supw
\left|nR^\ast(\w)-n\bar{R}(\w)\right|
\n\\&=&\xi_n^{-1}\supw\left|
\sumi
\int_0^1 \left\{
Q_{\oplus}(\x_i,t)-Q^\ast_{\oplus,i}(\w,t)
\right\}^2dt
-\mathbb{E}\sumi
\int_0^1 \left\{
Q_{\oplus}(\x_i,t)-\wh{Q}_{\oplus,i}(\w,t)
\right\}^2dt
\right|
\n\\&=&\xi_n^{-1}\supw\left|
\sumi
\int_0^1 \left\{
Q_{\oplus}(\x_i,t)-Q^\ast_{\oplus,i}(\w,t)
\right\}^2dt
-\mathbb{E}\sumi
\int_0^1 \left\{
Q_{\oplus}(\x_i,t)-Q^\ast_{\oplus,i}(\w,t)
\right\}^2dt
\right.\n\\&&\left.
+\mathbb{E}\sumi
\int_0^1 \left\{
Q_{\oplus}(\x_i,t)-Q^\ast_{\oplus,i}(\w,t)
\right\}^2dt
-\mathbb{E}\sumi
\int_0^1 \left\{
Q_{\oplus}(\x_i,t)-\wh{Q}_{\oplus,i}(\w,t)
\right\}^2dt
\right|
\n\\&\le& \xi_n^{-1}\supw
\left|
\mathbb{E}\sumi
\int_0^1 \left[\left\{
Q_{\oplus}(\x_i,t)-Q^\ast_{\oplus,i}(\w,t)
\right\}^2
-\left\{
Q_{\oplus}(\x_i,t)-\wh{Q}_{\oplus,i}(\w,t)
\right\}^2\right]dt
\right|
\n\\&&+\xi_n^{-1}\supw
\left|
\sumi
\left(
\int_0^1 \left\{
Q_{\oplus}(\x_i,t)-Q^\ast_{\oplus,i}(\w,t)
\right\}^2dt
\right.\right.\n\\&&\left.\left.
\ \ -\mathbb{E}
\int_0^1 \left\{
Q_{\oplus}(\x_i,t)-m^\ast_{\oplus,i}(\w,t)
\right\}^2dt
\right)
\right|
\n\\&\equiv&
\mathcal{A}_{1n}+\mathcal{A}_{2n}.
\label{eq:Th1:5}
\ee

For the first term of \eqref{eq:Th1:5}, we note that
\bse
&&
\mathbb{E}\sumi
\int_0^1 \left[\left\{
Q_{\oplus}(\x_i,t)-Q^\ast_{\oplus,i}(\w,t)
\right\}^2
-\left\{
Q_{\oplus}(\x_i,t)-\wh{Q}_{\oplus,i}(\w,t)
\right\}^2\right]dt
\\&=&\mathbb{E}\sumi
\int_0^1\left[
\left\{\wh Q_{\oplus}(\w,t)-Q_{\oplus}^\ast(\w,t)\right\}
\left\{\wh Q_{\oplus}(\w,t)+Q_{\oplus}^\ast(\w,t)
-2Q_{\oplus}(\x,t)\right\}\right]dt
\\&=& \mathbb{E}\sumi
 \int_0^1\Big[
-2Q_{\oplus}(\x,t)
\left\{\wh Q_{\oplus}(\w,t)-Q_{\oplus}^\ast(\w,t)\right\}
\\&&
+\left\{\wh Q_{\oplus}(\w,t)-Q_{\oplus}^\ast(\w,t)\right\}
\left\{\wh Q_{\oplus}(\w,t)+Q_{\oplus}^\ast(\w,t)\right\}
\Big]dt
\\&=& \mathbb{E}\sumi
\int_0^1\Big[
-2Q_{\oplus}(\x,t)
\left\{\wh Q_{\oplus}(\w,t)-Q_{\oplus}^\ast(\w,t)\right\}
\\&&
+\left\{\wh Q_{\oplus}(\w,t)-Q_{\oplus}^\ast(\w,t)\right\}^2
+2 Q_{\oplus}^\ast(\w,t)
\left\{\wh Q_{\oplus}(\w,t)-Q_{\oplus}^\ast(\w,t)\right\}
\Big]dt
\\&=&O(nc_n),
\ese
where the last equality is due to the
Assumption \ref{C:1} and \ref{C:2}. This leads to
$\mathcal{A}_{1n}= O_p(\xi_n^{-1}n c_n)
=o_p(1)$.
For the second term of \eqref{eq:Th1:5},
we have
\bse
&&
\supw
\left|
\sumi\left(
\int_0^1 \left\{
Q_{\oplus}(\x_i,t)-Q^\ast_{\oplus,i}(\w,t)
\right\}^2dt
-\mathbb{E}
\int_0^1 \left\{
Q_{\oplus}(\x_i,t)-Q^\ast_{\oplus,i}(\w,t)
\right\}^2dt
\right)
\right|
\\&=&
\supw
\left|
\sumi
\int_0^1 \left\{
Q_{\oplus}(\x_i,t)
-\sums
w_s Q_{\oplus}^\ast(\x_i^{(s)},t)
\right\}^2dt
\right.\\&&\left.
\ \
-\mathbb{E}\sumi\left[
\int_0^1 \left\{
Q_{\oplus}(\x_i,t)
-\sums w_s Q_{\oplus}^\ast(\x_i^{(s)},t)
\right\}^2dt\right]
\right|
\\&=&
\supw
\left|
\sumi
\int_0^1 \left[
\sums w_s\left\{
Q_{\oplus}(\x_i,t)
-m_{\oplus}^\ast(\x_i^{(s)},t)
\right\}
\right]^2dt
\right.\\&&\left.
\ \
-\mathbb{E}\sumi\left(
\int_0^1
\left[
\sums w_s
\left\{ Q_{\oplus}(\x_i,t)
- Q_{\oplus}^\ast(\x_i^{(s)},t)
\right\}
\right]^2dt\right)
\right|
\\&=&
\supw
\left|
\sumi
\int_0^1 \left[
\sum_{s_1=1}^S
\sum_{s_2=1}^S
w_{s_1} w_{s_2}
\left\{
Q_{\oplus}(\x_i,t)
-Q_{\oplus}^\ast(\x_i^{(s_1)},t)
\right\}
\left\{
Q_{\oplus}(\x_i,t)
-Q_{\oplus}^\ast(\x_i^{(s_2)},t)
\right\}
\right]
dt
\right.\\&&\left.
\ \
-\mathbb{E}\sumi\left\{
\int_0^1
\left[
\sum_{s_1=1}^S
\sum_{s_2=1}^S
w_{s_1} w_{s_2}
\left\{
Q_{\oplus}(\x_i,t)
-Q_{\oplus}^\ast(\x_i^{(s_1)},t)
\right\}
\left\{
Q_{\oplus}(\x_i,t)
-Q_{\oplus}^\ast(\x_i^{(s_2)},t)
\right\}
\right]
dt\right\}
\right|
\\&\le&
\sum_{s_1=1}^S
\sum_{s_2=1}^S
\left| \sumi
\left[
\int_0^1
\left\{
Q_{\oplus}(\x_i,t)
-Q_{\oplus}^\ast(\x_i^{(s_1)},t)
\right\}
\left\{
Q_{\oplus}(\x_i,t)
-Q_{\oplus}^\ast(\x_i^{(s_2)},t)
\right\}
dt
\right.\right.\\&&\left.\left.
\ \
-\mathbb{E}
\int_0^1
\left\{
Q_{\oplus}(\x_i,t)
-Q_{\oplus}^\ast(\x_i^{(s_1)},t)
\right\}
\left\{
Q_{\oplus}(\x_i,t)
-Q_{\oplus}^\ast(\x_i^{(s_2)},t)
\right\}
dt
\right]\right|
\\&=&O_p(n^{1/2}S^2),
\ese
where the last step is because
the $\mathcal{L}^2$-Wasserstein space
is
equivalent
to the subset of
$\mathcal{L}^2[0,1]$ formed by
quantile function on $[0,1]$, and
then the central limit theorem
can be applied.
By the Assumption \ref{C:3},
we have $\mathcal{A}_{2n}=o_p(1)$.
By above results and \eqref{eq:Th1:5},
we conclude that
\eqref{eq:Th1:3} holds.

Below we show \eqref{eq:Th1:4}.
We observe that
\bse
&&a_1(\w)\\
&=&\text{CV}_K^\ast(\w)-nR^\ast(\w)
\\&=&
\sum_{k=1}^K\sum_{j=1}^J
d_W^2\left\{Y_{(k-1)J+j},
\wh{m}_{\oplus,(k-1)J+j}^{[-k]}(\w)
\right\}
-\sumi
\int_0^1 \left\{
Q_{\oplus}(\x_i,t)-Y_i^{-1}(t)
\right\}^2dt
\\&&
-\sumi
\int_0^1 \left\{
Q_{\oplus}(\x_i,t)-Q^\ast_{\oplus,i}(\w,t)
\right\}^2dt
\\&=&
\sum_{k=1}^K\sum_{j=1}^J
\int_0^1
\left\{Y_{(k-1)J+j}^{-1}(t)-
\wh{m}_{\oplus,(k-1)J+j}^{[-k]}(\w,t)
\right\}^2dt
-\sumi
\int_0^1 \left\{
Q_{\oplus}(\x_i,t)-Y_i^{-1}(t)
\right\}^2dt
\\&&-\sumi
\int_0^1 \left\{
Q_{\oplus}(\x_i,t)-Q^\ast_{\oplus,i}(\w,t)
\right\}^2dt
\\&=&
\sum_{k=1}^K\sum_{j=1}^J
\int_0^1
\left[
\left\{
Y_{(k-1)J+j}^{-1}(t)
-Q_{\oplus,(k-1)J+j}(\x,t)\right\}
+
\left\{Q_{\oplus,(k-1)J+j}(\x,t)-
Q_{\oplus,(k-1)J+j}^\ast(\w,t)\right\}
\right.\\&&\left.
+\left\{Q_{\oplus,(k-1)J+j}^\ast(\w,t)-\wh{Q}_{\oplus,(k-1)J+j}(\w,t)\right\}
+\left\{\wh{Q}_{\oplus,(k-1)J+j}(\w,t)
-\wh{Q}_{\oplus,(k-1)J+j}^{[-k]}(\w,t)\right\}
\right]^2dt
\\&&
-\sumi
\int_0^1 \left\{
Q_{\oplus}(\x_i,t)-Y_i^{-1}(t)
\right\}^2dt
-\sumi
\int_0^1 \left\{
Q_{\oplus}(\x_i,t)-Q^\ast_{\oplus,i}(\w,t)
\right\}^2dt
\\&=&
\sum_{k=1}^K\sum_{j=1}^J\int_0^1
\left\{\wh{Q}_{\oplus,(k-1)J+j}(\w,t)
-\wh{Q}_{\oplus,(k-1)J+j}^{[-k]}(\w,t)\right\}^2dt
\\&&+2
\sum_{k=1}^K\sum_{j=1}^J
\int_0^1
\left\{
Y_{(k-1)J+j}^{-1}(t)
-\wh{Q}_{\oplus,(k-1)J+j}(\w,t)\right\}
\left\{\wh{Q}_{\oplus,(k-1)J+j}(\w,t)
-\wh{Q}_{\oplus,(k-1)J+j}^{[-k]}(\w,t)\right\}dt
\\&&+2\sum_{k=1}^K\sum_{j=1}^J\int_0^1
\left\{
Y_{(k-1)J+j}^{-1}(t)
-Q_{\oplus,(k-1)J+j}(\w,t)\right\}
\left\{Q_{\oplus,(k-1)J+j}(\x,t)-
Q_{\oplus,(k-1)J+j}^\ast(\w,t)\right\}dt
\\&&+2\sum_{k=1}^K\sum_{j=1}^J\int_0^1
\left\{
Y_{(k-1)J+j}^{-1}(t)
-Q_{\oplus,(k-1)J+j}(\w,t)\right\}
\left\{Q_{\oplus,(k-1)J+j}^\ast(\w,t)-\wh{Q}_{\oplus,(k-1)J+j}(\w,t)\right\}
dt
\\&&+2\sum_{k=1}^K\sum_{j=1}^J
\int_0^1
\left\{Q_{\oplus,(k-1)J+j}(\x,t)-
Q_{\oplus,(k-1)J+j}^\ast(\w,t)\right\}
\\&& \ \ \times
\left\{Q_{\oplus,(k-1)J+j}^\ast(\w,t)-\wh{Q}_{\oplus,(k-1)J+j}(\w,t)\right\}
dt
\\&&+
\sum_{k=1}^K\sum_{j=1}^J\int_0^1
\left\{Q_{\oplus,(k-1)J+j}^\ast(\w,t)
-\wh{Q}_{\oplus,(k-1)J+j}(\w,t)\right\}^2
dt
\\&&
+\left[\sum_{k=1}^K\sum_{j=1}^J
\int_0^1
\left\{
Y_{(k-1)J+j}^{-1}(t)
-Q_{\oplus,(k-1)J+j}(\x,t)\right\}^2dt
-\sumi
\int_0^1 \left\{
Q_{\oplus}(\x_i,t)-Y_i^{-1}(t)
\right\}^2dt\right]
\\&&+\left[
\sum_{k=1}^K\sum_{j=1}^J
\int_0^1
\left\{Q_{\oplus,(k-1)J+j}(\x,t)-
Q_{\oplus,(k-1)J+j}^\ast(\w,t)\right\}dt
\right.\\&&\left. \ \
-\sumi
\int_0^1 \left\{
Q_{\oplus}(\x_i,t)-Q^\ast_{\oplus,i}(\w,t)
\right\}^2dt
\right]
\n\\&=&
\mathcal{B}_{1n}+\mathcal{B}_{2n},
\ese
where
\bse
\mathcal{B}_{1n}
&=&
\sum_{k=1}^K\sum_{j=1}^J\int_0^1
\left\{\wh{Q}_{\oplus,(k-1)J+j}(\w,t)
-\wh{Q}_{\oplus,(k-1)J+j}^{[-k]}(\w,t)\right\}^2dt
\\&&
+2
\sum_{k=1}^K\sum_{j=1}^J
\int_0^1
\left\{
Y_{(k-1)J+j}^{-1}(t)
-\wh{Q}_{\oplus,(k-1)J+j}(\w,t)\right\}
\\&&\ \ \times
\left\{\wh{Q}_{\oplus,(k-1)J+j}(\w,t)
-\wh{Q}_{\oplus,(k-1)J+j}^{[-k]}(\w,t)\right\}dt,
\ese
and
\bse
\mathcal{B}_{2n}
&=&
2\sumi\int_0^1
\left\{
Y_i^{-1}(t)
-Q_{\oplus,i}(\w,t)\right\}
\left\{Q_{\oplus}(\x_i,t)-
Q_{\oplus,i}^\ast(\w,t)\right\}dt
\\&&
+2\sumi \int_0^1
\left\{
Y_i^{-1}(t)
-Q_{\oplus,i}(\w,t)\right\}
\left\{Q_{\oplus,i}^\ast(\w,t)-\wh{Q}_{\oplus,i}(\w,t)\right\}
dt
\\&&+2\sumi
\int_0^1
\left\{Q_{\oplus}(\x_i,t)-
Q_{\oplus,i}^\ast(\w,t)\right\}
\left\{Q_{\oplus,i}^\ast(\w,t)-\wh{Q}_{\oplus,i}(\w,t)\right\}
dt
\\&&+
\sumi\int_0^1
\left\{Q_{\oplus,i}^\ast(\w,t)
-\wh{Q}_{\oplus,i}(\w,t)\right\}^2
dt.
\ese
For $
\mathcal{B}_{2n}
$,
by the central limit theorem
and Assumptions \ref{C:1}
and \ref{C:2} with $Q_{\oplus}(\x_i,t)-
Q_{\oplus,i}^\ast(\w,t)=O_p(1)$,
it is seen that
\be
\mathcal{B}_{2n}
&=&O_p(n^{1/2})+O_p(n c_n)+
O_p(nc_n^2)
\n\\&=&O_p(n^{1/2})+O_p(n c_n).
\label{eq:Th1:6}
\ee
For $
\mathcal{B}_{1n}
$, we have
\bse
&&\mathcal{B}_{1n}
\\&=&
\sum_{k=1}^K\sum_{j=1}^J\int_0^1
\left\{\wh{Q}_{\oplus,(k-1)J+j}(\w,t)
-\wh{Q}_{\oplus,(k-1)J+j}^{[-k]}(\w,t)\right\}^2dt
\\&&+2
\sum_{k=1}^K\sum_{j=1}^J
\int_0^1
\left\{
Y_{(k-1)J+j}^{-1}(t)
-Q_{\oplus,(k-1)J+j}^\ast(\w,t)
+Q_{\oplus,(k-1)J+j}^\ast(\w,t)
-\wh{Q}_{\oplus,(k-1)J+j}(\w,t)\right\}
\\&&\ \ \ \times
\left\{\wh{Q}_{\oplus,(k-1)J+j}(\w,t)
-\wh{Q}_{\oplus,(k-1)J+j}^{[-k]}(\w,t)\right\}dt
\\&=&
\sum_{k=1}^K\sum_{j=1}^J\int_0^1
\left\{\wh{Q}_{\oplus,(k-1)J+j}(\w,t)
-\wh{Q}_{\oplus,(k-1)J+j}^{[-k]}(\w,t)\right\}^2dt
\\&&+2
\sum_{k=1}^K\sum_{j=1}^J
\int_0^1
\left\{
Y_{(k-1)J+j}^{-1}(t)
-Q_{\oplus,(k-1)J+j}^\ast(\w,t)
\right\}
\left\{\wh{Q}_{\oplus,(k-1)J+j}(\w,t)
-\wh{Q}_{\oplus,(k-1)J+j}^{[-k]}(\w,t)\right\}dt
\\&&+2
\sum_{k=1}^K\sum_{j=1}^J
\int_0^1
\left\{Q_{\oplus,(k-1)J+j}^\ast(\w,t)
-\wh{Q}_{\oplus,(k-1)J+j}(\w,t)\right\}
\\&& \ \ \times
\left\{\wh{Q}_{\oplus,(k-1)J+j}(\w,t)
-\wh{Q}_{\oplus,(k-1)J+j}^{[-k]}(\w,t)\right\}dt
\\&\le&
\sum_{k=1}^K\sum_{j=1}^J\int_0^1
\left\{\wh{Q}_{\oplus,(k-1)J+j}(\w,t)
-\wh{Q}_{\oplus,(k-1)J+j}^{[-k]}(\w,t)\right\}^2dt
\\&&+2
\sum_{k=1}^K\sum_{j=1}^J
\int_0^1
\left\{
Y_{(k-1)J+j}^{-1}(t)
-Q_{\oplus,(k-1)J+j}^\ast(\w,t)
\right\}
\left\{\wh{Q}_{\oplus,(k-1)J+j}(\w,t)
-\wh{Q}_{\oplus,(k-1)J+j}^{[-k]}(\w,t)\right\}dt
\\&&+
\sum_{k=1}^K\sum_{j=1}^J
\int_0^1
\left\{Q_{\oplus,(k-1)J+j}^\ast(\w,t)
-\wh{Q}_{\oplus,(k-1)J+j}(\w,t)\right\}^2dt
\\&& +\sum_{k=1}^K\sum_{j=1}^J
\int_0^1
\left\{\wh{Q}_{\oplus,(k-1)J+j}(\w,t)
-\wh{Q}_{\oplus,(k-1)J+j}^{[-k]}(\w,t)\right\}^2
dt.
\ese
Note that
from the formula of \eqref{eq:Rstar} and
Assumption \ref{C:2}, we have $n^{-1}\xi_n=O_p(1)$,
and combining with Assumption \ref{C:Jfold}, we obtain
\bse
&&
\left\{\wh{Q}_{\oplus,(k-1)J+j}(\w,t)
-\wh{Q}_{\oplus,(k-1)J+j}^{[-k]}(\w,t)\right\}^2
\\&=&
\left[
\sums w_s
\left\{
\wh{Q}_{\oplus}(\x_{(k-1)J+j}^{(s)},t)
-
\wh{Q}_{\oplus}^{[-k]}(\x_{(k-1)J+j}^{(s)},t)
\right\}
\right]^2
\\&=&o_p(n^{-1}\xi_n)
\\&=&o_p(1).
\ese
Based on above result
and Assumption \ref{C:Jfold}, we have
\be
\mathcal{B}_{1n} &=&
o_p(\xi_n)+o_p(\xi_n)+
O_p(nc_n^2)
+o_p(\xi_n)
\n\\&=& o_p(\xi_n)+
O_p(nc_n^2)
. \label{eq:Th1:7}
\ee

Therefore, by \eqref{eq:Th1:6}
and \eqref{eq:Th1:7}, we have
$
a_1(\w)=
O_p(n^{1/2})+O_p(nc_n)
+o_p(\xi_n).
$
Based on
$R^\ast(\w)=O_p(1)$ deduced
from Assumption \ref{C:2},
we can see that \eqref{eq:Th1:4} holds.
This completes the proof of \eqref{eq:Th1:1}.

For \eqref{eq:Th1:2}, by
the proof of \eqref{eq:Th1:5},
it follows that
\bse
n\bar{R}(\w)-nR(\w)
&=&
n\bar{R}(\w)-nR^\ast(w)+nR^\ast(w)-nR(\w)
\\&=&nR^\ast(w)-n R(\w)+ O_p(nc_n)+O_p(n^{1/2}S^2).
\ese
Note that $\wh{m}_{\oplus,i}^{[-i]}(\w,t)$
is the estimator
of $m_{\oplus}(\x_i,t)$
without using the $i$th observation,
so it shares the same limits as
$\wh{m}_{\oplus,i}(\w,t)$.
Thus, we also have
$nR^\ast(w)-n R(\w)=
nO(n^{-1/(2(\beta-1))})+O_p(n^{1/2}S^2)$,
which entails that
$\bar{R}(\w)-R(\w)=O_p(c_n)+O_p(n^{-1/2}S^2)
=o_p(1)$, by Assumption \ref{C:3}.
Then,
this establishes the
\eqref{eq:Th1:2},
which completes the proof.
\qed

\noindent{
{\bf Proof of Theorem \ref{th:2}.}}
For the sake of simplicity in notation,
let $\tau=\sum_{s \in \mathcal{C}} w_s$
and
$\wh{\tau}=\sum_{s \in \mathcal{C}} \wh{w}_s$.
We next to show that $\wh{\tau} \rightarrow 1$ in
probability. Further,
let $\boldsymbol\lambda$ be also a
weight vector of $S$ dimension with
$\lambda_s=0$ for $s \in \mathcal{C}$
and $\lambda_s=w_s /(1-\tau)$
for $s \notin \mathcal{C}$.
From the proof of \eqref{eq:Th1:4},
we know that
\bse
\text{CV}_K^\ast(\w)/n
-R^\ast(\w)
&=&O_p(n^{-1/2})+O_p(c_n)
+o_p(\xi_n/n)\\
&=&O_p(n^{-1/2})+O_p(c_n)
+o_p(1).
\ese
Note that the above result also holds
by replacing weight $\w$ with $\wh\w$. That is,
\be
\text{CV}_K^\ast(\wh\w)/n
-R^\ast(\wh\w)
&=&O_p(n^{-1/2})+O_p(c_n)
+o_p(1).
\label{eq:Th2:1}
\ee
Note that
\be
R^\ast(\w)
&=& \frac{1}{n}\sumi
d_W^2\{m_{\oplus}(\x_i),m^\ast_{\oplus,i}(\w)\}
\n\\&=&\frac{1}{n}\sumi
\int_0^1 \left\{
Q_{\oplus}(\x_i,t)-Q^\ast_{\oplus,i}(\w,t)
\right\}^2dt
\n\\&=&
\frac{1}{n}\sumi
\int_0^1
\left[
\sums w_s
\left\{
Q_{\oplus}(\x_i,t)-
Q_{\oplus}^\ast(\x_i^{(s)},t)
\right\}
\right]^2
dt
\n\\&=&
\frac{1}{n}\sumi
\int_0^1
\left[
\sum_{s \notin \mathcal{C}} w_s
\left\{Q_{\oplus}(\x_i,t)-
Q_{\oplus}^\ast(\x_i^{(s)},t)
\right\}
\right]^2 dt
\n\\&=&(1-\tau)^2
\frac{1}{n}\sumi
\int_0^1
\left[
\sum_{s \notin \mathcal{C}}
\frac{w_s}{1-\tau}
\left\{Q_{\oplus}(\x_i,t)-
Q_{\oplus}^\ast(\x_i^{(s)},t)
\right\}
\right]^2
dt
\n\\&=&(1-\tau)^2
\frac{1}{n}\sumi
\int_0^1
\left[
\sums
\lambda_s
\left\{
Q_{\oplus}(\x_i,t)-
Q_{\oplus}^\ast(\x_i^{(s)},t)
\right\}
\right]^2
dt
\n\\&=&(1-\tau)^2 R^\ast(\boldsymbol{\lambda}).
 \label{eq:Th2:3}
\ee
It is also easy to see that the
above result holds
by replacing weight $\w$ with $\wh\w$,
and together with \eqref{eq:Th2:1} and
\eqref{eq:Th2:3}, we have
\be
\text{CV}_K^\ast(\wh\w)/n
&=&(1-\wh\tau)^2 R^\ast(\wh{\boldsymbol{\lambda}})
+O_p(n^{-1/2})+O_p(c_n)
+o_p(1),
 \label{eq:Th2:4}
\ee
where
$\wh{\lambda}_s=\widehat{w}_s /(1-\widehat{\tau})$
when $ s \notin \mathcal{C}$
and $\wh{\lambda}_s=0$
when $s \in \mathcal{C}$.
Let $\wt\w$ be a weight vector
satisfying $\sum_{s\in\mathcal{C}}\wt{w}_s=1$.
For any correctly specified model
$s \in \mathcal{C}$, we have
$
m_{\oplus}(\x_i,t)
-m_{\oplus}^\ast(\x_i^{(s)},t)
=0,
$
from which
and formula of \eqref{eq:Th2:3},
we have $R^\ast(\wt\w)=0$.
Then,
it follows that, from \eqref{eq:Th2:1},
\be
\text{CV}_K^\ast(\wt\w)/n
&=&O_p(n^{-1/2})+O_p(c_n)
+o_p(1).  \label{eq:Th2:5}
\ee
Recall that the fact $\wh\w$ minimizes
$\text{CV}_K^\ast(\w)$,
and combining \eqref{eq:Th2:4} and  \eqref{eq:Th2:5},
we have
\bse
(1-\wh\tau)^2 R^\ast(\wh{\boldsymbol{\lambda}})
+O_p(n^{-1/2})+O_p(c_n)
+o_p(1)
&\le&
\text{CV}_K^\ast(\wt\w)/n
=O_p(n^{-1/2})+O_p(c_n)
+o_p(1),
\ese
from which, we have
\bse
(1-\wh\tau)^2
\inf_{\w\in \cal{W}_S}R^\ast(\w)
&\le&
O_p(n^{-1/2})+O_p(c_n)
+o_p(1).
\ese
Thus,
based on above result and Assumption \ref{C:weight},
we conclude that
$\wh\tau\overset{\Pr}\rightarrow 1$.
The proof is completed.
\qed

\clearpage

\begin{table}[!h]
\centering
\caption{
Monte Carlo results for
averaged risk.
The sample size $n$
equals to $100$, $200$ and $300$, respectively.
Standard deviations are given in
parentheses.
In each setting,
the method with the best
performance is marked in bold.
}\label{tab:rho5R9}
\renewcommand{\arraystretch}{1.2}
\begin{tabular}{ccccccccccc}
\toprule[1.3pt]
      & {$n$}
	  & $100$
      & $200$
      & $300$ \\
\midrule[1.3pt]
& $\text{CV}$     &  {\bf 0.0709} (0.0273)
&  {\bf 0.0324} (0.0154)  &  {\bf0.0217} (0.0121) \\
& $\text{sAIC}$   &  0.0756 (0.0249)  &  0.0381 (0.0170)  &  0.0244 (0.0119) \\
& $\text{sBIC}$   &  0.0786 (0.0248)  &  0.0389 (0.0177)  &  0.0248 (0.0120) \\
& $\text{EW}$     &  0.0759 (0.0259)  &  0.0372 (0.0166)  &  0.0245 (0.0120) \\
& $\text{Oracle}$ &  0.0612 (0.0238)  &  0.0268 (0.0144)  &  0.0190 (0.0118) \\
& $\text{Full}$   &  0.1489 (0.0450)  &  0.0780 (0.0248)  &  0.0525 (0.0165) \\
& $\text{Ridge}$  &  0.1114 (0.0361)  &  0.0525 (0.0200)  &  0.0337 (0.0163) \\
& $\text{AIC}$    &  0.1387 (0.0456)  &  0.0696 (0.0258)  &  0.0460 (0.0169) \\
& $\text{BIC}$    &  0.2520 (0.1073)  &  0.0946 (0.0667)  &  0.0417 (0.0141) \\
\bottomrule[1.3pt]
\end{tabular}
\end{table}

\begin{table}[!h]
\centering
\caption{The sum of weights
assigned to correct candidate models.}
\label{tab:weight}
\renewcommand{\arraystretch}{1.2}
\begin{tabular}{ccccccccccc}
\toprule[1.3pt]
$\text{Scenario}$
&
&\multicolumn{3}{c}{Sum}\\
\midrule[1.3pt]
\multirow{1}*{$n=100$}
&& & 0.9211 &  \\
\hline
\multirow{1}*{$n=200$}
&& &0.9723 & \\
\hline
\multirow{1}*{$n=300$}
&& & 0.9939 &  \\
\bottomrule[1.3pt]
\end{tabular}
\end{table}


%

\clearpage

\bibliographystyle{agsm}
\bibliography{frechet_MA_0901}

\end{document}